\documentclass[a4paper,11pt]{article}
\pdfoutput=1 % if your are submitting a pdflatex (i.e. if you have
\usepackage{jcappub} % for details on the use of the package, please
\usepackage{txfonts} % damit ich keinen Augenkrebs bekomme
\usepackage{cleveref}
\usepackage{enumerate}
\usepackage{tensor}

\usepackage{color}

\def\karl#1{{\color{black}{ #1}}}
\def\karlo#1{{\color{black}{ #1}}}
\newcommand{\ltsima}{$\; \buildrel < \over \sim \;$}
\newcommand{\lsim}{\lower.5ex\hbox{\ltsima}}
\newcommand{\gtsima}{$\; \buildrel > \over \sim \;$}
\newcommand{\gsim}{\lower.5ex\hbox{\gtsima}}

\newcommand{\dd}{\mathrm{d}}

\title{\karl{\boldmath Information geometry in cosmological inference problems}}

\author[a]{Eileen Giesel,}
\author[a,b,c,d]{Robert Reischke,}
\author[a]{Bj\"orn Malte Sch\"afer,}
\author[a]{and Dominic Chia}

\affiliation[a]{Astronomisches Rechen-Institut, Zentrum f{\"u}r Astronomie der Universit{\"a}t Heidelberg, Philosophenweg 12, 69120 Heidelberg, Germany}
\affiliation[b]{Institut f\"ur Kernphysik, Karlsruher Institut f\"ur Technologie, 76344 Eggenstein-Leopoldshafen, Germany}
\affiliation[c]{Department of Physics, Technion, Haifa 32000, Israel}
\affiliation[d]{Department of Natural Sciences, The Open University of Israel, 1 University Road, P.O. Box 808, Ra'anana 4353701, Israel}

\emailAdd{e.giesel@stud.uni-heidelberg.de, r.reischke@technion.ac.il}

\abstract{\karl{ 
Statistical inference often involves models which are non-linear in the parameters and which therefore typically exhibit non-Gaussian posterior distributions. These non-Gaussianities can be prominent especially when data is limited or not constraining enough. Many computational and analytical tools exist that can deal with non-Gaussian distributions, and empirical Gaussianisation transforms can be constructed that can reduce the amount of non-Gaussianity in a distribution. In this work, we employ methods from information geometry, which considers a set of probability distributions for some given model to be a manifold with a metric Riemannian structure, given by the Fisher information. In this framework we study the differential geometrical meaning of non-Gaussianities in a higher order Fisher approximation, and their respective transformation behaviour under re-parameterisation, which corresponds to a chart transition on the statistical manifold. While weak non-Gaussianities vanish in normal coordinates in a first order approximation, one can in general not find transformations that discard non-Gaussianities globally. As a topical application in cosmology we consider the likelihood of the supernovae distance-redshift relation for the parameter pair ($\Omega_{\mathrm{m0}}$, $w$). We show that the corresponding manifold is non-flat and demonstrate the connection between confidence intervals and geodesic length, determine the curvature of that likelihood and quantify degeneracies by means of Lie-derivatives.}
\
\\
\
}

\begin{document}
\maketitle
\flushbottom

% --- section: introductin --- %
\section{Introduction}
Any predictive physical model contains a set of finitely many parameters which need to be determined by experiment. These parameters can be both physical parameters of interest, or nuisance parameters taking care of e.g. instrumental systematics which will eventually be marginalised. \karl{Physical parameters can be, for example, the matter density $\Omega_\mathrm{m}$, the Hubble-Lema{\^i}tre constant $h$, the dark energy equation of state $w$, or the sum of the neutrino masses $\sum m_\nu$, while nuisance parameters could appear in a weak lensing measurement as the amplitude of the intrinsic alignment contamination. Depending on the construction of the model, the measurement method and the data volume, a model can be constrained to a certain domain in parameter space where the model might or might not be well approximated by a linear relationship. This situation is especially severe in cosmology for the following two reasons: $(i)$ Data from cosmological observations is inherently limited, due to cosmic variance on very large scales and astrophysical effects on small scales. $(ii)$ There is usually a hierarchy in the model parameters and it is not possible to test individual parameters with a designed experiment independently. Then, if the data is not constraining enough to restrict the parameter space to a small neighbourhood, one can not expect the likelihood to be of Gaussian shape because a linearisation of the model over the allowed parameter space is not applicable.}

\karl{Both in a Bayesian as well as in a frequentist framework the central object is the joint probability density function $p(\boldsymbol{x},\boldsymbol{\theta})$ of the parameters, $\boldsymbol{\theta}$, and the data $\boldsymbol{x}$. The true model is then suspected to be the one that could have brought about the data with the highest likelihood, i.e. the choice of the parameters that is able to maximise $p(\boldsymbol{x},\boldsymbol{\theta})$. Bayesian inference will interpret $p(\boldsymbol{\theta}|\boldsymbol{x})$ rather in terms of confidence in a particular set of parameters given a single realisation of the data. Cosmological inference usually follows this route since, by nature, the data is only available in a single realisation, casting a Frequentist interpretation questionable.}

Values of model parameters are given with confidence intervals \karl{which describe the degree of uncertainty of the derived parameter value given the errors of the particular experimental technique. These intervals are usually determined with two main formalisms: For forecasting, where the best fit values of the parameters are assumed to be known and one often assumes likelihoods close to Gaussian shapes, mainly the Fisher-matrix analysis \citep[e.g.][]{tegmark_karhunen-loeve_1997} or higher order schemes \citep{Sellentinetal, sellentin_fast_2015} are employed.} Instead, when facing real data or for more accurate confidence contours without an assumption of (near) Gaussianity or of single-modality of the likelihood, one relies on Monte-Carlo Markov-chain (MCMC) methods \citep{skilling_nested_2006, akeret_cosmohammer:_2013, feroz_multinest:_2011, audren_monte_2013, goodman_ensemble_2010}. Approximating a likelihood as being Gaussian has tremendous computational advantages compared to MCMC-techniques, and requires only a small number of calls of the likelihood for finite differencing, at the expense of not representing the likelihood faithfully enough.

The origin of non-Gaussian features of a likelihood are inherent non-linearities in the dependence of the model on its parameters or a non-Gaussian error process for the individual data points, giving rise as well to constraints and degeneracies that vary as a function of the fiducial model \citep{schafer_describing_2016}. If the data is well-constraining, these non-linear relationships can be linearised, which render the $\chi^2$-functional parabolic and cast the likelihood to be Gaussian. When a model is extended to include new parameters, the corresponding likelihood covers a larger fraction of the parameter space and a linearisation of the model might not be applicable, in which case the likelihood veers again away from the Gaussian shape. This effect can be counteracted by accumulating more data, by collecting data with smaller errors, or by combining different measurement methods with the potential of breaking degeneracies.

Whether it is possible to re-parameterise a model such that an otherwise non-Gaussian likelihood assumes a Gaussian shape is an interesting question and has a clear positive answer in one dimension. For multivariate distributions, this can only be done approximatively \citep{box_analysis_1964, schuhmann_gaussianization_2016}, making the likelihood accessible to arguments reserved to Gaussian distributions, for instance their unbiasedness and their fulfilment of the Cram{\'e}r-Rao-bound, but also the perfect decoupling of all degeneracies by transforming the parameter space into the eigensystem of the Gaussian's covariance matrix. Whether a perfect Gaussianisation can in principle be achieved is an interesting question on its own and it can be answered using the tools of information geometry. In information geometry, the Fisher-matrix (or, the inverse parameter covariance) takes on the role of the metric on a statistical manifold \citep{amari_information_2016} and can be derived from a statistical divergence defined in an axiomatic way.
In this setting the parameters of the statistical model are merely a choice of coordinates {and one should be able to change the parameterisation by chart transition maps in an invertible and differentiable way}. {In this context,} Gaussian likelihoods correspond to flat manifolds {(although there might be unfortunate parameter choices where they appear to be non-Gaussian)} while {actual} non-Gaussian ones induce a non-trivial geometry and a deviation from flatness, quantified by a curvature tensor. \karl{Hence, using the concepts of information geometry leads to an understanding of non-Gaussianities as inherent geometrical properties of a non-flat manifold which is defined by a statistical model. In the case of non-flat manifolds it is in general not possible to find a coordinate change to obtain a constant Fisher information corresponding to a Gaussian likelihood in the parameters. Additionally, by employing additional methods of differential geometry it could still be possible to find for instance isometries of the Fisher information and consequently integral curves in parameter space along which the metric is constant, leading to a geometric interpretation of the degeneracies of a model and indicating parameter choices where Gaussianity of the likelihood is established.}

In this work we will apply the concepts of information geometry to a typical statistical model encountered in cosmology. As a working example we will choose the statistical manifold defined by supernova observations \citep{riess_observational_1998, perlmutter_discovery_1998, perlmutter_measurements_1999, riess_type_2004, riess_new_2007} and investigate its differential geometric properties in the context of a flat $w$CDM-cosmology, and restrict the inference on the matter density $\Omega_m$ and a constant dark energy equation of state $w$. Furthermore, we will discuss Gaussianisation transformations, as well as the relation between the DALI-approximation \citep{sellentin_fast_2015} and the Gram-Charlier series with geometric properties of statistical manifolds in the limit of weak non-Gaussianities, providing relationships between these fundamental descriptions of non-Gaussian distributions. Of course, scientists do not only intend to measure the model parameters of a single model as accurate as possible, but also look out for new phenomena beyond the accepted model. In that, we will take the point of view that a model class is already selected, for instance with concepts of Bayesian evidence or through a strong theoretical argument, and we are asking how well different parameter choices within this model class are compatible with data.

\karl{The structure of our paper reflects the scientific motivation to apply methods from information geometry to an inference problem in cosmology: We will recapitulate the basic concepts of information geometry in \cref{sec:information_geometry}. \Cref{sec:Gaussianization} will be devoted to the geometrical interpretation of Gaussianity and its transformation properties, where we establish links between curvature of a statistical manifold and alternative descriptions of non-Gaussianity of multivariate distributions. In \cref{sec:example} we will apply these ideas to the example of constraints on a $w$CDM-cosmology from type-Ia supernovae as a topical example of a bivariate non-Gaussian distributions. Specifically, we compute the Ricci-curvature for the manifold formed by the two parameters $\Omega_m$ and $w$, and investigate degeneracies on that manifold with Lie-derivatives. Finally, we summarise our results in \cref{sec:summary}.}

% --- section: concepts --- %
\section{Concepts of information geometry}
\label{sec:information_geometry}
In this section we will briefly summarise the basic concepts of information geometry \citep{amari_information_2016}. We will introduce the statistical manifold, $M$, under consideration, the divergence which itself induces a metric $g$ on $M$. Additional structure is then provided by choosing a particular linear connection $\nabla$ such that concepts like curvature tensors and curvature scalars can be introduced on the triple $(M,g,\nabla)$. \karl{The manifold M will turn out to be of Riemannian type equipped with a positive definite metric $g$ with a Euclidean signature in contrast to relativity where a pseudo-Riemannian manifold with a metric of Lorentzian signature is considered and the classification of vectors in time-, space- or light-like is done on the basis of their norm: This is due to the fact that in relativity the concept of hyperbolic spacetimes is central along with a geometric notion of causality, both of which is irrelevant to statistical manifolds.}

% ---  --- %
\subsection{Statistical manifold}
At the heart of inference is a statistical model which can be described as a set
\begin{equation}\label{eq:set_statistical_model}
M = \{p(\boldsymbol{x},\boldsymbol{\theta})\},
\end{equation}
where $\boldsymbol{\theta}$ are the model parameters, $\boldsymbol{x}$ the data and $p$ is a probability or probability density, in the case of a continuum of parameters. This set has the structure of a $d$-dimensional topological manifold \cite{amari_information_2016}, that is a paracompact Hausdorff topological ($\tau$, a suitable topology) space $(M,\tau)$ such that $\forall p\in M $ there exists an open neighbourhood $U$ with a homeomorphism $U\to U^\prime\subseteq \mathbb{R}^d$. Here, the dimensionality is given by the dimensions of the parameter space. In physics the homeomorphism is usually referred to as a coordinate system. As the statistical model is described by a set of parameters $\boldsymbol{\theta}$, they can be seen as the coordinate system parameterising the manifold. We will assume here that the manifold is smooth, i.e. that all chart transition maps in its atlas are $C^\infty$.

% ---  --- %
\subsection{Divergences and invariant metric}
Distances between points on $M$ are described by divergences, which quantify the dissimilarity of the distributions associated with every point of the likelihood. For $p,q\in M$ we write $\boldsymbol{\theta}_i$ as the corresponding coordinate. A divergence is then given by
\begin{equation}\label{eq:divergence_general}
D[p:q] = D[\boldsymbol{\theta}_p,\boldsymbol{\theta}_q], 
\end{equation}
where the following criteria must hold:
\begin{enumerate}
\item $D[p:q] \geq 0$.
\item $D[p:q] = 0$ if and only if $p = q$.
\item $D[p:q]$ can be Taylor expanded in the local coordinate system if $p$ and $q$ are sufficiently close to each other:
\begin{equation*}
D[\boldsymbol{\theta}_p,\boldsymbol{\theta}_p+\dd \boldsymbol{\theta}]=\frac{1}{2}{g}_{ij}(\boldsymbol{\theta}_p)\dd\theta^i\dd\theta^j,
\end{equation*} 
where the matrix $\boldsymbol{g}$ is positive definite. 
\end{enumerate}
Clearly, the squared infinitesimal distance can thus be written as
\begin{equation}\label{eq:squared_infi_distance}
\dd s^2 =2 D[\boldsymbol{\theta}_p,\boldsymbol{\theta}_p+\dd \boldsymbol{\theta}],
\end{equation}
providing a Riemannian structure on $M$. The natural divergence between two points $p$ and $q$ can be derived by demanding 
\begin{equation}\label{eq:invariance}
D[p:q] \geq D^\prime [p:q],
\end{equation}
where $D^\prime$ denotes the divergence associated to another random variable $\boldsymbol{y} = \phi(\boldsymbol{x})$. The statistic $\boldsymbol{y}$ is called sufficient if the equality in (\ref{eq:invariance}) holds. Clearly any one-to-one mapping provides a sufficient statistic. 

The divergence between two points $p$ and $q$ measures the mutual information between the two distributions at those two points. An invariant information measure can be introduced \cite{1963Morimoto} called the $f$-divergence: 
\begin{equation}\label{eq:f_divergence}
D_f[\boldsymbol{\theta}_p,\boldsymbol{\theta}_q] = 
\int \dd \boldsymbol{x}\:p(\boldsymbol{x}) \:f\left(\frac{q(\boldsymbol{x})}{p(\boldsymbol{x})}\right),
\end{equation}
with a differentiable and convex function satisfying $f(1)=0$. This divergence can be shown to be invariant. A typical example for an $f$-divergence is the Kullback-Leibler divergence for which $f(x) =-\log(x)$. Using the properties of a divergence it is clear that the positive definite matrix $\boldsymbol{g}$ of the $f$-divergence provides a natural invariant metric on $M$. It can be seen easily that any $f$-divergence{, provided $f^\prime (1) = 0$ and $f^{\prime\prime}(1) = 1$, i.e. if $f$ is standard,} yields the same Riemannian metric which is the Fisher information matrix:
\begin{equation}
\begin{split}
D_f[\boldsymbol{\theta}_p,\boldsymbol{\theta}_p, +\mathrm{d}\boldsymbol{\theta}] 
= & \ \int \mathrm{d}\boldsymbol{x}\: p(\boldsymbol{x}|\boldsymbol{\theta}_p)\left[\frac{f^{\prime\prime}(1)}{2}\frac{\mathrm{d}p}{\mathrm{d}\theta^\mu}\frac{\mathrm{d}p}{\mathrm{d}\theta^\nu}\bigg|_{\mathrm{d}\boldsymbol{\theta} = 0} \frac{\mathrm{d}\theta^\mu\mathrm{d}\theta^\nu}{p(\boldsymbol{x}|\boldsymbol{\theta}_p)^2}\right], \\
= & \ \frac{1}{2}\int\mathrm{d}\boldsymbol{x}\: p(\boldsymbol{x}|\boldsymbol{\theta}_p)\left(\frac{\mathrm{d}p}{\mathrm{d}\theta^\mu}\frac{1}{p(\boldsymbol{x}|\boldsymbol{\theta}_p)}\frac{\mathrm{d}p}{\mathrm{d}\theta^\nu}\frac{1}{p(\boldsymbol{x}|\boldsymbol{\theta}_p)}\right) \\ 
= & \ \frac{1}{2}\left\langle \frac{\partial \mathrm{ln}p(\boldsymbol{x}|\boldsymbol{\theta}_p)}{\partial\theta^\mu}\frac{\partial \mathrm{ln}p(\boldsymbol{x}|\boldsymbol{\theta}_p)}{\partial\theta^\nu}\right\rangle \mathrm{d}\theta^\mu \mathrm{d}\theta^\nu,
\end{split}
\end{equation}
Comparing this result to the third property of the divergence we will therefore write the metric as:

\begin{equation}\label{eq:fisher_information_metric}
g_{ij} =  
\left\langle\frac{\partial \log p(\boldsymbol{x},\boldsymbol{\theta})}{\partial\boldsymbol{\theta}^i}\frac{\partial \log p(\boldsymbol{x},\boldsymbol{\theta})}{\partial\boldsymbol{\theta}^j}\right\rangle = 
\int\dd \boldsymbol{x}\;p(\boldsymbol{x},\boldsymbol{\theta})  \partial_i \log p(\boldsymbol{x},\boldsymbol{\theta}) \partial_j \log p(\boldsymbol{x},\boldsymbol{\theta}).
\end{equation}
More importantly this metric is unique up to a constant factor. In \cite{amari_information_2016} the geometrical inner product $g_{ij} = \langle \boldsymbol{e}_i, \boldsymbol{e}_j \rangle$ is identified with the statistical Fisher information such that the tangent vectors $\boldsymbol{e}_i$ can be related to the score functions $\boldsymbol{e}_i \sim \partial \log p(\boldsymbol{x},\boldsymbol{\theta})/\partial\boldsymbol{\theta}^i$ as derivative of the logarithmic likelihood. As a side remark the Riemannian structure on $M$ and hence the positive definiteness of the metric is also vital for the validity of the Cram{\'e}r-Rao inequality which states that the inverse Fisher information evaluated at the likelihoods best fit is a lower bound for the parameter covariance \cite[e.g.][]{tegmark_karhunen-loeve_1997}.

% ---  --- %
\subsection{Connection and cubic tensor}\label{sec:connection_and_cubic_tensor}
The statistical manifold (\ref{eq:set_statistical_model}) is equipped with a metric~(\ref{eq:fisher_information_metric}) and thus assumes the structure of a Riemannian manifold $(M,\boldsymbol{g})$. We also introduce an affine connection $\nabla$ which allows for the notion of parallel transport, geodesics and curvature on the object $(M,\boldsymbol{g},\nabla)$. It should be noted that the connection $\nabla$ is completely general. However, we will assume parallel transport to not affect the magnitude of vectors, thus restricting ourself to metric connections. {Furthermore, the connection is assumed to be symmetric \cite{amari_information_2016} what corresponds to the absence of torsion. If metricity is presumed one can even give an argument for the connection to be symmetric as follows: In a chart representation with $\lbrace\boldsymbol{e}_{i}\rbrace = \lbrace\partial/ \partial\theta^{i}\rbrace$ as coordinate basis of the tangent space $T_p M$ the connection coefficient functions $\Gamma^{i}_{j \, k}$ are in general defined as $\nabla_{i}\,\boldsymbol{e}_j = \Gamma^{k}_{j \, i} \boldsymbol{e}_k$ \cite[][p.250]{Nakahara_Geometry_2003}, and in case of metricity this definition even simplifies to the partial derivative $\partial_i\,\boldsymbol{e}_j = \Gamma^{k}_{j \, i} \boldsymbol{e}_k $ \cite[][p.63]{Hobson_GR_2006}. For the definition of the basis vectors $\boldsymbol{e}_j$ in information geometry this becomes $\partial_{i}\,\boldsymbol{e}_j = \Gamma^{k}_{j \, i} \boldsymbol{e}_k = \partial^2 \log p(\boldsymbol{x},\boldsymbol{\theta})/\partial\boldsymbol{\theta}^i \,\partial\boldsymbol{\theta}^j $ what is clearly symmetric in the lower two indices $i$ and $j$.} Thus the connection is given by the torsion free Levi-Civita connection. In this latter case geodesic lines trace curves of minimal distances between two points on $M$, if they are affinely parameterised.

Indeed, the Levi-Civita connection is the unique torsion free affine connection preserving the norm of a vector under parallel transport. However, if we write the inner product of two vectors $\boldsymbol{u}$, $\boldsymbol{v}$, which are parallel transported, in the following way \citep{amari_information_2016}:
\begin{equation}\label{eq:norm_vector_parallel_transport}
\langle \boldsymbol{u},\boldsymbol{v}\rangle = 
g(\boldsymbol{u},\boldsymbol{v}) = 
\langle \nabla \boldsymbol{u},\nabla^*\boldsymbol{v}\rangle,
\end{equation}
if the two connections $\nabla$ and $\nabla^*$ are chosen such that they do not change the inner product they are said to be dually coupled. In the case that $\nabla = \nabla^*$ the Levi-Civita connection is recovered. For the general case, one can show the following relation \citep{amari_information_2016}:
\begin{equation}\label{eq:dual_transport}
\Gamma_{ijk} = \partial_i g_{jk} -\Gamma_{ikj}^*, 
\end{equation}
where {$\Gamma_{ijk} =g_{il}\,\Gamma^{l}_{ij}$ and $\Gamma_{ijk}^*=g_{il}\,\Gamma^{*\,l}_{ij}$} are the connection coefficients associated with $\nabla$ and $\nabla^*$ respectively {with indices lowered by metric contraction}. Defining the tensor
\begin{equation}\label{eq:cubic_tensor}
T_{ijk} = \Gamma^*_{ijk} - \Gamma_{ijk}, 
\end{equation}
{and comparing expression (\ref{eq:dual_transport}) to the first derivative of the metric expressed in terms of the Levi-Civita connection
\begin{equation}\label{eq:LC_transport}
\partial_i g_{jk} =\Gamma^{\mathrm{LC}}_{ijk} + \Gamma^{\mathrm{LC}}_{ikj}, 
\end{equation}}
the connection coefficients are obtained as
\begin{equation}\label{eq:connection_coefficients}
\Gamma_{ijk} = 
\Gamma^{\mathrm{LC}}_{ijk} -\frac{1}{2}T_{ijk},
\quad 
\Gamma^*_{ijk} = \Gamma^{\mathrm{LC}}_{ijk} +\frac{1}{2}T_{ijk},
\end{equation}
with $\Gamma^{\mathrm{LC}}_{ijk}$ being the connection coefficients of the Levi-Civita connection, namely the Christoffel symbols. The invariant cubic tensor $T_{ijk}$ derived from an $f$-divergence is given by
\begin{equation}\label{eq:cubic_tensor}
T_{ijk} = 
\alpha \left\langle\frac{\partial \log p(\boldsymbol{x},\boldsymbol{\theta})}{\partial\boldsymbol{\theta}^i}\frac{\partial \log p(\boldsymbol{x},\boldsymbol{\theta})}{\partial\boldsymbol{\theta}^j}\frac{\partial \log p(\boldsymbol{x},\boldsymbol{\theta})}{\partial\boldsymbol{\theta}^k}\right\rangle,
\end{equation}
with $\alpha = 2f^{\prime\prime\prime}(1) + 3$. In this sense the statistical manifold (\ref{eq:set_statistical_model}) can also be thought of as given by the triple $(M,\boldsymbol{g},\boldsymbol{T})$. {However, in special cases like for a likelihood of the form
\begin{equation}\label{eq:example_likelihood}
p\left(\boldsymbol{x} \vert \, \boldsymbol{\mu}\left(\boldsymbol{\theta}\right)\right) = 
\frac{1}{\sqrt{\left( 2 \pi \right)^n \text{det}\boldsymbol{C}}}\, \exp^{\left(-\frac{1}{2} \boldsymbol{X}^T \, \boldsymbol{C}^{-1} \, \boldsymbol{X} \right)},
\end{equation} 
with a constant data covariance $ \boldsymbol{C}$ and data vector $\boldsymbol{X} \coloneqq \boldsymbol{x} - \boldsymbol{\mu}\left(\boldsymbol{\theta}\right)$ the cubic tensor vanishes. In this case, the score functions are proportional to $\boldsymbol{X}$ and odd moments of a (multivariate) Gaussian vanish due to Isserlis' Theorem \cite{isserlis_theorem}. This will turn out to be important for the examples considered in \cref{sec:Gaussianization}.}

% ---  --- %
\subsection{Integration on manifolds}
An invariant volume element, $\mathrm{d}\Omega_M$, on a manifold is a $d$-form: 
\begin{equation}
\mathrm{d}\Omega_M = \sqrt{\mathrm{det}\boldsymbol{g}}\;  \dd\theta^1\wedge\dots\wedge\dd\theta^d.
\end{equation}
The factor $\sqrt{\mathrm{det}\boldsymbol{g}}$ ensures that the volume element is invariant and can be directly interpreted: Consider for example the normalisation condition of a Gaussian distribution with zero mean, which does not restrict the generality of the argument,
\begin{equation}
\int \mathrm{d}^{\,d}\theta\:\sqrt{\frac{\mathrm{det}\boldsymbol{F}}{(2\pi)^d}}\exp\left(-\frac{1}{2}\boldsymbol{\theta}^\mathrm{T}\boldsymbol{F}\boldsymbol{\theta}\right) = 1.
\end{equation}
Comparing the two expressions show that effectively the co-volume factor $\sqrt{\mathrm{det}\boldsymbol{F}}$ is merged with the Euclidean volume element $d^d\theta$ to form the invariant volume element $\mathrm{d}\Omega_M$, such that a reparameterisation does not change the normalisation. Furthermore, $\boldsymbol{F}$ becomes the Fisher metric which is constant in case of Gaussian likelihoods. It should be noted that the averaging not only removes the {explicit} dependence on the data, but also ensures that a canonical volume form on the statistical manifold is given.

% --- section: Gaussianisation --- %
\section{Gaussianisation from an information geometric viewpoint}
\label{sec:Gaussianization}
As outlined in \Cref{sec:information_geometry}, the invariant infinitesimal distance between two neighbouring points on the statistical manifold $(M,F)$ is given by $\mathrm{d}s^2= F_{ij}\mathrm{d}\theta^i\mathrm{d}\theta^j$, where the metric is derived from the divergence, \cref{eq:divergence_general}. Here, a specific coordinate system, or parameter set, $\{\theta_i \}$ has been chosen. A Gaussian distribution with respect to the parameters would correspond to the case where $\boldsymbol{F}$ is independent from the parameters, $\boldsymbol{\theta}$. This, however, is a parameter dependent statement as the components of $\boldsymbol{F}$ transform as:
\begin{equation}
\label{eq:transformation_law_fisher_matrix}
\tensor{F}{^\prime_{ij}}(\boldsymbol{\theta}^\prime) = 
\tensor{J}{^a_i}(\boldsymbol{\theta}^\prime) \tensor{J}{^b_j}(\boldsymbol{\theta}^\prime)\tensor{F}{_a_b}(\boldsymbol{\theta}), 
\end{equation}
with the Jacobian $J^a_{~b}\coloneqq \partial \theta^a /\partial \theta^{\prime b}$. If we are to find any transformation \cref{eq:transformation_law_fisher_matrix} which leads to a globally parameter independent Fisher matrix, the manifold $(M,F)$ would be flat and there would be a global Gaussianisation transformation. Using this argument in the opposite way, even a likelihood described by a flat manifold can show non-Gaussian structure, depending on the chosen coordinate system. In particular, for an originally uncorrelated Gaussian distribution with unit variance one could generate non-Gaussianities through the transformation
\begin{equation}
\begin{split}
\tensor{F}{^\prime_{ij,k}}(\boldsymbol{\theta}^\prime) = \tensor{J}{_{ak,i}}(\boldsymbol{\theta}^\prime)\tensor{J}{^a _j}(\boldsymbol{\theta}^\prime) + \tensor{J}{^a_{j,k}}(\boldsymbol{\theta}^\prime)\tensor{J}{_{ai}}(\boldsymbol{\theta}^\prime)\;.
\end{split}
\end{equation}
Here, we denote the partial derivative as $\partial_a f \equiv f_{,a}$. The commonly used Fisher matrix approach to forecast the sensitivity of future experiments by virtue of the Cram{\'e}r-Rao bound assumes that the pair $(M,F)$ is a flat manifold, if not extended to deal with non-Gaussianities  \citep{Sellentinetal, sellentin_fast_2015, schafer_describing_2016}. Going beyond the Fisher approximation thus includes terms which might be attributed to the non-vanishing curvature of the statistical manifold, and we aim to derive relations between this geometric point of view and conventional descriptions of non-Gaussianity.

% ---  --- %
\subsection{Weak non-Gaussianitites in the Gram-Charlier-limit: {A first approach}}
\label{sec:wng_GCL}
\karl{In this section we turn to weak non-Gaussianities, i.e. the cumulants have to satisfy $\kappa_n / \left(\sqrt{\sigma^2}\right)^n \ll 1$ (in one dimension and  accordingly in the multivariate case). Furthermore, we will use the phrase vanish with different meanings. If something vanishes to some order (with respect to an expansion scheme) it means that the quantity is zero up to this order. That is, higher orders generate non-vanishing terms. Vanish for weak non-Gaussianities uses the definition just given. If the non-Gaussianities are not assumed to be weak, certain manipulations to arrive at the result would not have been well defined.}

Going back to the invariant infinitesimal element $\mathrm{d}s^2= F_{ij}\mathrm{d}\theta^i\mathrm{d}\theta^j$ the distance between any two points $P$ and $Q$ along a curve $c(\lambda)$ is 
\begin{equation}
\label{eq:distance_finite}
D(P,Q) \propto
\left( \int_{\lambda(P)}^{\lambda(Q)} \mathrm{d}\lambda\:\sqrt{F[\dot{c}(\lambda),\dot{c}(\lambda)]}\right)^2,
\end{equation}
where the dot refers to the derivative with respect to the curve parameter $\lambda$. For $D(P,Q)$ to be the shortest distance, $c(\lambda)$ has to be a geodesic of the metric $\boldsymbol{F}$. \Cref{eq:distance_finite} is of particular insight when considering a Gaussian (in the data) likelihood, because in this case $D(P,Q) \propto \Delta\chi^2(P,Q)$, thus we can stipulate that the likelihood for the parameters to be $P(\boldsymbol{\theta}) \propto \exp(-D(P,Q)/2)$, such that $\boldsymbol{\theta}$ is the image of $Q$ under some chart. The point $P$ is just for reference and can be absorbed in the proportionality constant, reflecting the fact that only differences in $\chi^2$ are of any relevance. {Since a likelihood with Gaussian distributed data is considered, we can assume that the cubic tensor (\ref{eq:cubic_tensor}) vanishes as argued in \cref{sec:connection_and_cubic_tensor}. Thus we can assume the Levi-Civita connection in our further calculations instead of the dually coupled ones.}

Going back to the definition of the distance between two points on the manifold, \cref{eq:distance_finite}, one can use the expansion of the metric, $\boldsymbol{F}$ to write the distance in terms of Gaussian and non-Gaussian contributions
\begin{equation}
D(P,Q) \approx 
\left( \int_{\lambda (P)}^{\lambda(Q)} \mathrm{d}\lambda\:\sqrt{\left(\tensor{F}{^0_{ab}} +\tensor{F}{^0_{ab,g}}\Delta\theta^g +\frac{1}{2}\tensor{F}{^0_{ab,gd}}\Delta\theta^g\Delta\theta^d \right) \dot\theta^\alpha\dot\theta^b}  \right)^2,
\end{equation}
where the superscript $0$ denotes evaluation at the point $P$ (for instance the best-fit point) and $\Delta\theta^\alpha \coloneqq \theta^\alpha - \theta(P)^\alpha$. 
The latter equation can be rearranged using the inverse metric and thus to split everything into Gaussian and perturbatively non-Gaussian parts:
\begin{equation}
\label{eq:weak_NG_integrand}
D(P,Q) \approx  
\left( 
\int_{\lambda (P)}^{\lambda(Q)} \mathrm{d}\lambda\:\sqrt{\tensor{F}{^0_{ib}}\dot\theta^a\dot\theta^b} \left(\delta{^i_a} + \tensor{F}{^{0ir}}\tensor{F}{^0_{ar,g}}\Delta\theta^g +\frac{1}{2}\tensor{F}{^{0ir}}\tensor{F}{^0_{ar,gd}}\Delta\theta^g\Delta\theta^d \right)^{1/2}
\right)^2.
\end{equation}
This expression is of course very similar to a multidimensional Gram-Charlier series, which expands a distribution around its Gaussian part, assuming that higher order cumulants are small (compared to the variance). 
We now choose normal coordinates at the point $\boldsymbol{\theta}_0$. In these coordinates geodesics are again Euclidean straight lines and $\dot\theta^a = a^a$ {while the connection coefficient functions $\Gamma^{\mathrm{LC}}_{ijk}$ vanish locally \cite{lovelock_rund_tensors}}. \Cref{eq:weak_NG_integrand} can now be expanded further and then integrated trivially to find:
\begin{equation}
D(P,Q) \approx  \left(\tensor{F}{^0_{ab}}\Delta\theta^a\Delta\theta^b +\frac{1}{2}\tensor{F}{^0_{ab,g}}\Delta\theta^a\Delta\theta^b\Delta\theta^g +\frac{1}{6}\tensor{F}{^0_{ab,gd}}\Delta\theta^a\Delta\theta^b\Delta\theta^g \Delta\theta^d \right). 
\end{equation}
The remaining terms can be simplified further: the first term, {which just includes the first derivative of the metric and as such depends on the connection $\Gamma^{\mathrm{LC}}_{ijk}$ according to relation (\ref{eq:LC_transport})}, again vanishes in normal coordinates. For the second one, a bit more work is required: First we note that in normal coordinates we can write
\begin{equation}\label{eq:second_derivative_normal}
\tensor{F}{^0_{ab,g,d}} = 
-\frac{1}{3}\left(R_{agbd} - R_{adbg}\right).
\end{equation}
{Here $R_{agbd} = g_{ac} R^{c}_{~gbd}$ denotes the components of the Riemann curvature tensor which is here defined with respect to the Levi-Civita connection. Contracting the expression (\ref{eq:second_derivative_normal}) with $\Delta\theta^a\Delta\theta^b\Delta\theta^g \Delta\theta^d $ can be shown to vanish in a straightforward calculation due to the symmetries of the Riemann tensor (especially $R_{agbd} = -R_{gabd}$ and $R_{agbd} = - R_{agdb}$).} This shows that non-Gaussianities can only play a role at second order in a suitable chosen coordinate system. 

{The expansion scheme of the likelihoods' exponent in terms of the Riemannian distance is a first effective approach. One decisive advantage is that $D(P,Q)$ is positive definite by definition and allows for a first and fast estimate of the effect of non-Gaussianity for a non constant Fisher information. In fact it was shown \cite{amari_information_2016} that there is a close relation between a symmetrised Kullback-Leibler divergence and Riemannian distances. However, the full Kullback-Leibler divergence is asymmetric in general. Thus we now need to turn to a different expansion scheme, which also complies with information due to this asymmetry.}

% --- section:  --- %
\subsection{Curvature, Gaussianisation and the DALI expansion}\label{sec:curvature_gaussianization_dali}
There are alternative expansions of non-Gaussian likelihooods \citep{sellentin_fast_2015}: In particular the integrand of \cref{eq:distance_finite} can be expanded in terms of higher order derivatives of {the Fisher Information}. Put differently one can say that it is an expansion of the Fisher matrix around the best fit point $\boldsymbol{\theta}_0$. {However in \citep{sellentin_fast_2015} not the Fisher information but more generically the logarithmic likelihood $L$ itself is expanded around the best fit $\boldsymbol{\theta}_0$ and only afterwards one performs a data average over the expansion coefficients.} In particular one defines the flexion and the quarxion as:
\begin{equation}
\tensor{S}{^0_{abg}} =
\left\langle \tensor{L}{_{,abg}}\right\rangle\big|_{\boldsymbol{\theta}_0}, 
\quad 
\tensor{Q}{^0_{abgd}} =
\left\langle \tensor{L}{_{,abgd}}\right\rangle\big|_{\boldsymbol{\theta}_0},
\end{equation}
respectively. Expanding the likelihood this way, allows for the calculation of the Kullback-Leibler divergence relative to the fiducial point. The likelihood for the parameters can now be written as
\begin{equation}\label{eq:dali_expansion}
p(\boldsymbol{\theta}) \propto 
\exp\left[-\frac{1}{2}\tensor{F}{^0_{ab}}\Delta\theta^a\Delta\theta^b - \frac{1}{3!}\tensor{S}{^0_{abg}}\Delta\theta^a\Delta\theta^b \Delta\theta^g -\frac{1}{4!}\tensor{Q}{^0_{abgd}} \Delta\theta^a\Delta\theta^b \Delta\theta^g\Delta\theta^d
\right],
\end{equation}
{with a logarithmic likelihood expansion up to fourth order}. In this sense, it is an expansion of the distribution $p(\boldsymbol{\theta})$ relative to some fiducial distribution $p(\boldsymbol{\theta}_0)$. In particular the average over the data is again necessary to measure the divergence between the two distributions. Crucially, this expansion is different from the one presented in \cref{sec:wng_GCL} since it works directly on the level of the Kullback-Leibler divergence and is therefore not symmetric, that is the integrand in \cref{eq:f_divergence} is expanded before the averaging is carried out. Therefore, this expansion is not necessarily symmetric. In contrast the expansion in \cref{sec:wng_GCL} relies on the expansion of a Riemannian distance measure, which of course is symmetric. 

For a Gaussian sampling distribution, DALI expands the likelihood in terms of derivatives with respect to the mean or the covariance, depending which of them carries the parameter dependence. This ensures that the remaining distribution is still a proper probability distribution function. In particular the flexion and the quarxion contracted with sufficiently many $\Delta\theta^a$s can be shown to be always positive definite. Since both the flexion and the quarxion contain non-Gaussian information at third and fourth order about the likelihood a naive guess would be to relate them to the higher order cumulants of the expansion. Indeed, one can derive the following relations using a Gram-Charlier ansatz, i.e. for weak non-Gaussianities:\karl{
\begin{equation}\label{kumulant_relations}
\begin{split}
\kappa^a =& \ -\frac{1}{6}\tensor{S}{^0_{ijk}}\tensor{A}{^{ijka}}, \\
\kappa^{abc} =& \ -\tensor{S}{^0_{ijk}}F^{ia}F^{jb}F^{kc}, \\
\kappa^{ab} =& \ F^{ab}-\frac{1}{12}\tensor{Q}{^0_{ijkl}}\tensor{D}{^{ijklab}}, \\
\kappa^{abcg} =& \ -\tensor{Q}{^0_{ijkl}}F^{ia}F^{jb}F^{kc}F^{lg}.
\end{split}
\end{equation}
Here we have introduced the following abbreviations:
\begin{equation}
\begin{split}
A^{abcd} \coloneqq & \  F^{ab}F^{cd} + F^{ac}F^{bd} + F^{ad}F^{cb},\\
D^{abcdij} \coloneqq & \ F^{ac}F^{bi}F^{dj} + F^{ai}F^{bc}F^{dj} + F^{aj}F^{bi}F^{cd} + F^{ci}A^{abdj},
\end{split}
\end{equation}}
Thus all additional terms are proportional to the Fisher matrix and thus scale with different powers on $\sigma^{-2}$. {In appendix \ref{Multi_Gram} more details are given about the multivariate Gram-Charlier expansion and we sketch in more detail how the expressions (\ref{kumulant_relations}) can be derived from that.} An alternative way to arrive at these expressions is to use directly the Fa{\`a} di Bruno formula, relating flexion and quarxion directly to the moments.

We will now compare the quarxion and the flexion to differential geometric quantities on the statistical manifold. We first start from the definition of the flexion and rewrite it as
\begin{equation}
\label{eq:flexion}
S_{abc} = 
\tensor{\Gamma}{^{0\,,\mathrm{LC}}_{cab}} +
\tensor{\Gamma}{^{0\,,\mathrm{LC}}_{acb}} + 
\tensor{\Gamma}{^{0\,,\mathrm{LC}}_{bac}} + 
\frac{1}{2}\left\langle L_{,a}L_{,b}L_{,c}\right\rangle,
\end{equation}
where $\tensor{\Gamma}{^{0\,,\mathrm{LC}}_{cab}} = g_{ic}\tensor{\Gamma}{^{0i\,,\mathrm{LC}}_{ab}}$ are the Christoffel symbols, which are defined in terms of derivatives of the Fisher metric, and we evaluate everything again at the best fit point $\boldsymbol{\theta}_0$, where we can identify the last term as the cubic tensor. Thus, in normal coordinates, where the Christoffel symbols vanish, the flexion is completely sourced by the cubic tensor. {For a likelihood as discussed in relation (\ref{eq:example_likelihood}) in \cref{sec:connection_and_cubic_tensor} the cubic tensor, and hence the flexion vanishes completely in normal coordinates.} A similar exercise can be performed for the quarxion, where we seek a relationship with the second derivatives of the metric. In particular it can be shown that the quarxion vanishes in normal coordinates to first order in the second derivatives of the metric:
\begin{equation}
\label{eq:quarxion}
\begin{split}
Q_{ijkl} = & -\frac{1}{3}\left(
R_{ikjl} + R_{iljk} -R_{ikjl} +R_{lkji} + R_{klij} +R_{kjil}  - R_{iljk} -R_{klji} - R_{klij} +R_{ikjl}-R_{kjil} -R_{ikjl}
\right)\\
= & \ 0 + \mathcal{O}(L_{,ijk}),
\end{split}
\end{equation}
where we used the symmetry properties of the Riemann tensor, which can be expressed, in normal coordinates, in terms of second derivatives of the metric for the Levi-Civita connection. Both results, for the flexion and the quarxion, apply for the dependence of the covariance on the parameters as well as for the case where the mean depends on the parameters. $(L_{,ijk})$ refers to terms which contain proper third derivatives of the quantities carrying the parameter dependence. The results presented in this section only hold for weak non-Gaussianities, i.e. up to first order. However, the curvature does not depend on derivatives of third order. For definiteness let's assume a likelihood of the form \cref{eq:loglike_sn}, where the parameter dependence is only carried by $\boldsymbol{\mu}$ such that
\begin{equation}
\begin{split}
R_{ijkl}  = & \boldsymbol{\mu}^\mathrm{T}_{\,,ik}\boldsymbol{C}^{-1}\boldsymbol{\mu}_{,jl} - \boldsymbol{\mu}^\mathrm{T}_{\,,il}\boldsymbol{C}^{-1}\boldsymbol{\mu}_{,jk}.
\end{split}
\end{equation}
{Finally, using concepts from information geometry we found an interpretation of non-Gaussianities as inherent geometrical properties of a statistical manifold. These can in general only be remedied for a flat manifold by a non-linear coordinate transformation. For a non-flat manifold they only vanish in normal coordinates in first order approximation, so one can in principle not find a non-linear coordinate transformation to make them vanish completely. However, one could still search for isometries of the Fisher information as will be discussed in the next section, where we consider the example a likelihood on $\Omega_\mathrm{m0}$ and $w$ from supernova data.}

% --- section: example --- %
\section{An illustrative example: The Supernovae likelihood}\label{sec:example}
In this section we will discuss a simple cosmological example in the context of information geometry: distance measurements with supernova-observations. {We will investigate the geometrical properties of the manifold defined by the likelihood, especially geodesics and the Ricci scalar $R = g^{ik}R_{ik} = g^{ik}g^{jl}R_{ijkl}$, and investigate the Lie-derivatives to search for isometries of the Fisher information.}

\begin{figure}[h!]
\begin{center}
\includegraphics[keepaspectratio,height=9.7cm, width=9.7cm]{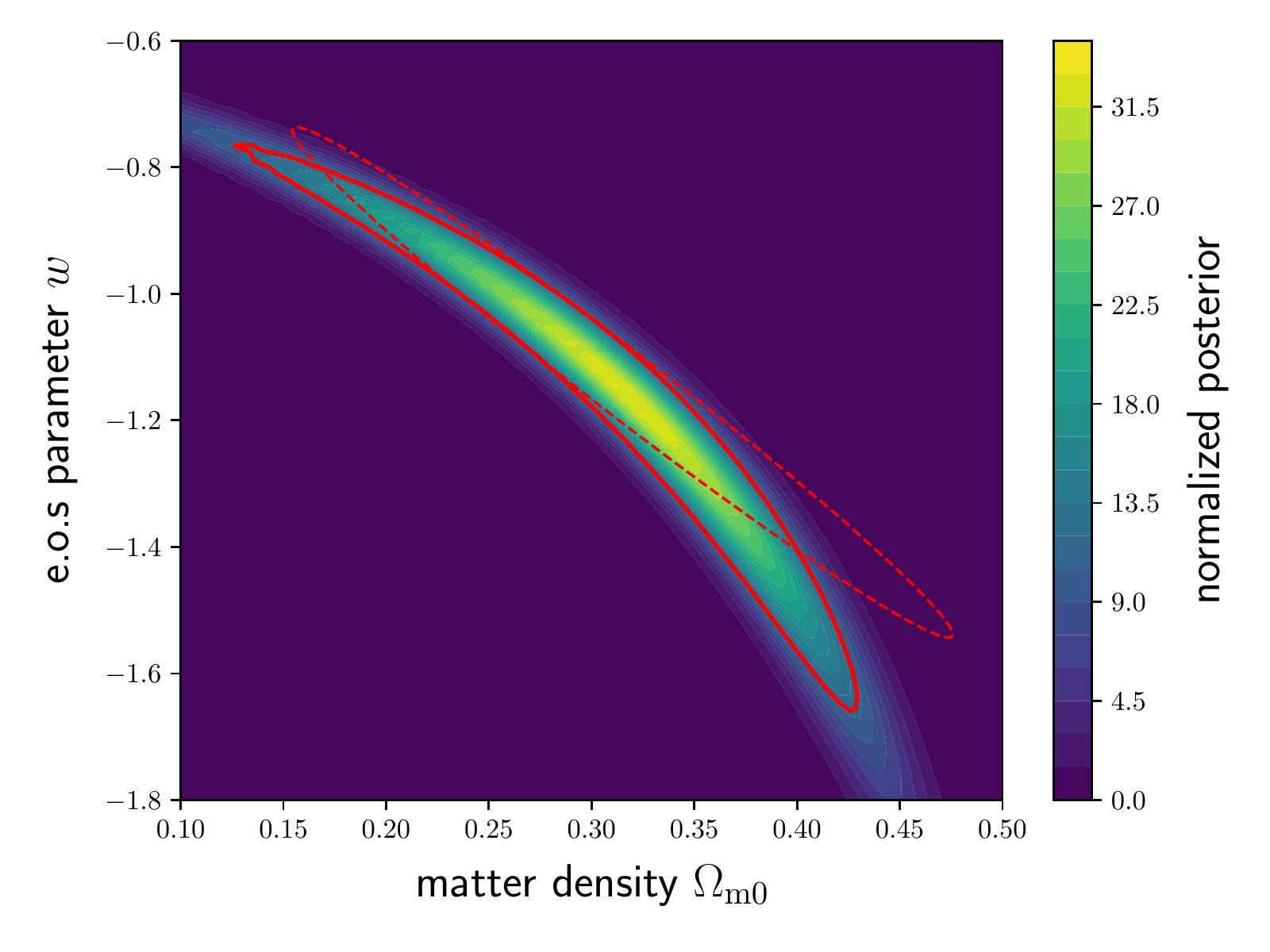}
\caption{\karl{Exact posterior, \cref{eq:loglike_sn}, and its Fisher approximation using \cref{eq:fisher_matrix_sn}. We show the 68$\;\%$ confidence interval (solid red) for the exact likelihood and its Fisher approximation (dashed red).}} 
\label{fig:likelihood}
\end{center}
\end{figure}

% ---  --- %
\subsection{The invariant metric}
As an example we consider the Gaussian likelihood for a supernovae measurement. It has the following simple form \cite{Sellentinetal}
\begin{equation}\label{eq:loglike_sn}
\mathcal{L}(D|\theta) \propto \exp(\boldsymbol{\mu}^T\boldsymbol{C}^{-1}\boldsymbol\mu),
\end{equation} 
with the vector
\begin{equation}\label{eq:mu_sn}
\boldsymbol{\mu} \coloneqq \boldsymbol{m} -\boldsymbol{m}_\mathrm{theory}.
\end{equation}
The two vectors $\boldsymbol{m}$ and $\boldsymbol{m}_\mathrm{theory}$ bundle the observed values for the distance modulus and the corresponding model prediction for a given set of parameters $\{\theta\}$ at $n$ redshifts $z_i$ respectively. The errors of different measurements are encoded in the  covariance matrix defined as $\boldsymbol{C}\coloneqq \langle\boldsymbol{\mu}\otimes\boldsymbol{\mu}\rangle$. Finally, the distance modulus is defined as
\begin{equation}\label{eq:distance_modulus}
m(z|\{\theta\}) = 5 \log d_\mathrm{lum}(z|\{\theta\}) +\mathrm{const},
\end{equation}
where the luminosity distance can be calculated from the background cosmology as
\begin{equation}\label{eq:luminosity_distance}
d_\mathrm{lum}(z|\{\theta\}) = c\int_0^z\frac{(1+z^\prime)\mathrm{d}z^\prime}{H(a(z^\prime|\{\theta\})}, 
\end{equation}
where the Hubble function $H(a)=\dot{a}/a$ is given by,
\begin{equation}\label{eq:Hubble_function}
\frac{H^2(a)}{H_0^2} = \frac{\Omega_\mathrm{m_0}}{a^{3}} + \frac{1-\Omega_\mathrm{m_0}}{a^{3(1+w)}},
\end{equation}
for a constant equation of state function $w$ \citep{2001IJMPD..10..213C, 2006APh....26..102L, 2008GReGr..40..329L}. We will assume the covariance in Eq. \eqref{eq:loglike_sn} to be diagonal and parameter independent:
\begin{equation}
\boldsymbol{C} = \mathrm{diag}(\sigma^2_1,...,\sigma^2_n).
\end{equation}
The corresponding Fisher matrix is easily derived to be
\begin{equation}\label{eq:fisher_matrix_sn}
F_{ab}(\{\theta\}) = \sum_{i=1}^n \frac{\partial_a m(z_i)}{\sigma_i}\frac{\partial_b m(z_i)}{\sigma_i}\bigg|_{\{\theta\}}.
\end{equation}

% ---  --- %
\subsection{Differential geometric quantities}\label{sec:differential_geometric_quantities}
We will now treat the supernova likelihood as a statistical manifold $(M,\boldsymbol{g})$ with the metric given by \cref{eq:fisher_matrix_sn}. {As discussed in \cref{sec:connection_and_cubic_tensor} the cubic tensor of the likelihood in our example, given by eq. (\ref{eq:loglike_sn}) vanishes, so we can assume the Levi-Civita connection.} \karl{For illustrative purposes we show the exact likelihood together with the the Fisher approximation in the $(\Omega_\mathrm{m0},w)$-plane in \cref{fig:likelihood}, illustrating deviations from Gaussianity caused by the nonlinear dependence on $\Omega_\mathrm{m0}$ and $w$. The solid red contour shows the exact 1-$\sigma$ contour, while the dashed red one corresponds to the Fisher approximation.}
Clearly, the Fisher approximation does not capture the complete shape of the likelihood. However, good agreement can be already achieved with the first two terms using the DALI approximation \citep{Sellentinetal}\karl{, i.e. the flexion and the quarxion. We have seen before that the former should vanish in normal coordinates, \cref{eq:flexion}. If the flexion would describe the exact likelihood already quite well one could deduce that the underlying statistical manifold should be flat. This already gives a hint that the statistical manifold is non-flat.}

\begin{figure}[h!]
\begin{center}
\includegraphics[keepaspectratio,height=9.7cm, width=9.7cm]{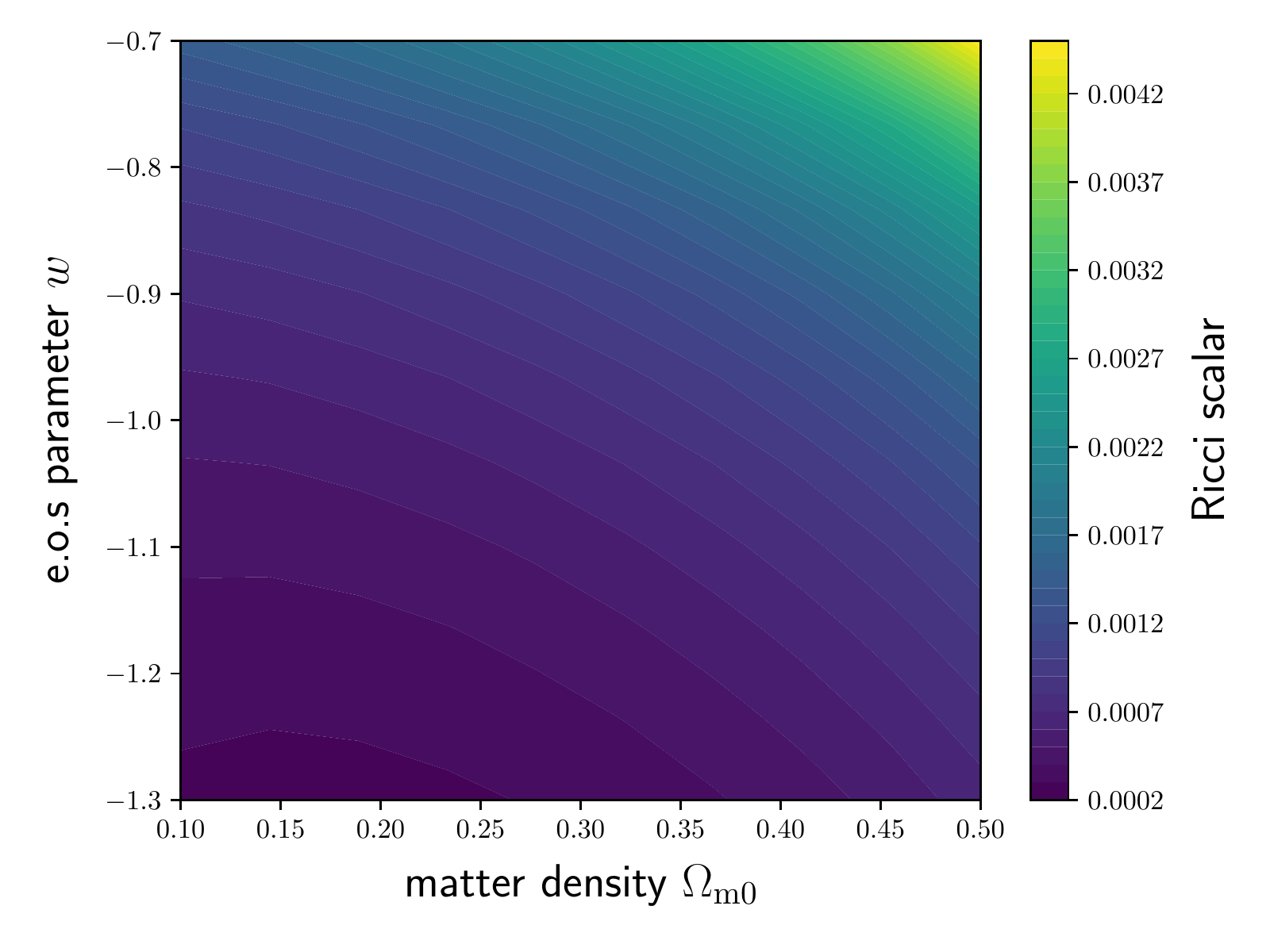}
\caption{The Ricci scalar, i.e. the curvature scalar $R = g^{ik}g^{jl}R_{ijkl}$, for the statistical manifold with the metric given by the supernova-measurement \cref{eq:fisher_matrix_sn}. Since $R$ is a scalar it is invariant under a change of coordinates and therefore provides a measure of the non-Gaussianity of the underlying statistical model irrespective of the choice of parameters.} 
\label{fig:ricci}
\end{center}
\end{figure}

\karl{To gain a better understanding of the geometrical properties of the statistical manifold we calculate the Ricci-curvature, \Cref{fig:ricci}, as a function of the coordinates $\Omega_\mathrm{m0}$ and $w$ as they would result from the supernova-measurement. Notably, the scalar curvature does not vanish, showing that the non-Gaussianity of the likelihood is inherent to the manifold and is not due to a pure, although physically motivated, choice of parameters. One can, however, see that the Ricci scalar is very small but non-vanishing and it increases with increasing $\Omega_\mathrm{m0}$ and $w_0$. The smallness of the Ricci scalar provides some important information about this parameter space: While the manifold is not-flat globally, $L\sim1/\sqrt{R}$ provides the size of the local neighbourhood in which the manifold can be considered approximately flat: This estimate is a direct consequence from the fact that curvature is derived from second derivatives of the metric. Recall that in this case $R$ has units $[\Omega_\mathrm{m0}w]^{-1}$. Therefore, we expect curvature effects to become important at scales $L$, which is in this case 10-100. This means that non-Gaussian effects become only important if we change $\Omega_\mathrm{m0}$ or $w_0$ by a similar amount. Clearly, \cref{fig:likelihood} shows non-Gaussian features, however, the smallness of the Ricci scalar only tells us that it is possible to find a new set of parameters in which the manifold looks flat over the extent of $L$. We emphasise that this argument reflects only geometric properties of the likelihood and does not use any physical constraint on the range of possible values that the parameters could take, for instance a bound $w\geq -1$.}

\karl{In general there is no straightforward equivalent of the decomposition of the Riemann tensor into a Ricci- and a Weyl-part as in relativity, where the Ricci curvature is linked to the energy-momentum tensor by the field equation and the Weyl tensor contains the propagating components of curvature. We can, however, make use of the fact that only in manifolds with four or more dimensions there can be Weyl curvature, such that in our two-dimensional case the Ricci curvature contains all information about the geometry of the manifold and that the Ricci scalar provides a straightforward measure of the total curvature. Only in four dimensions or more the Weyl tensor carries independent information about the curvature. In this case the Kretschmann scalar, $\mathcal{K} \equiv R_{ijkl}R^{ijkl}$, would be the proper object to study.}

\karl{Two-dimensional manifolds as the one considered in our example are conformally flat, too, such that the actual metric can be constructed from the Euclidean metric by a conformal transformation, but we leave it to a later study if that conformal transformation has implications for the construction of Gaussianisation transforms, for instance for slowly varying conformal factors. At the same time we would like to point out that due to the Euclidean signature of the metric there is no notion of null-geodesics and their conformal invariance.}

\begin{figure}[h!]
\begin{center}
\includegraphics[keepaspectratio,height=9.7cm, width=9.7cm]{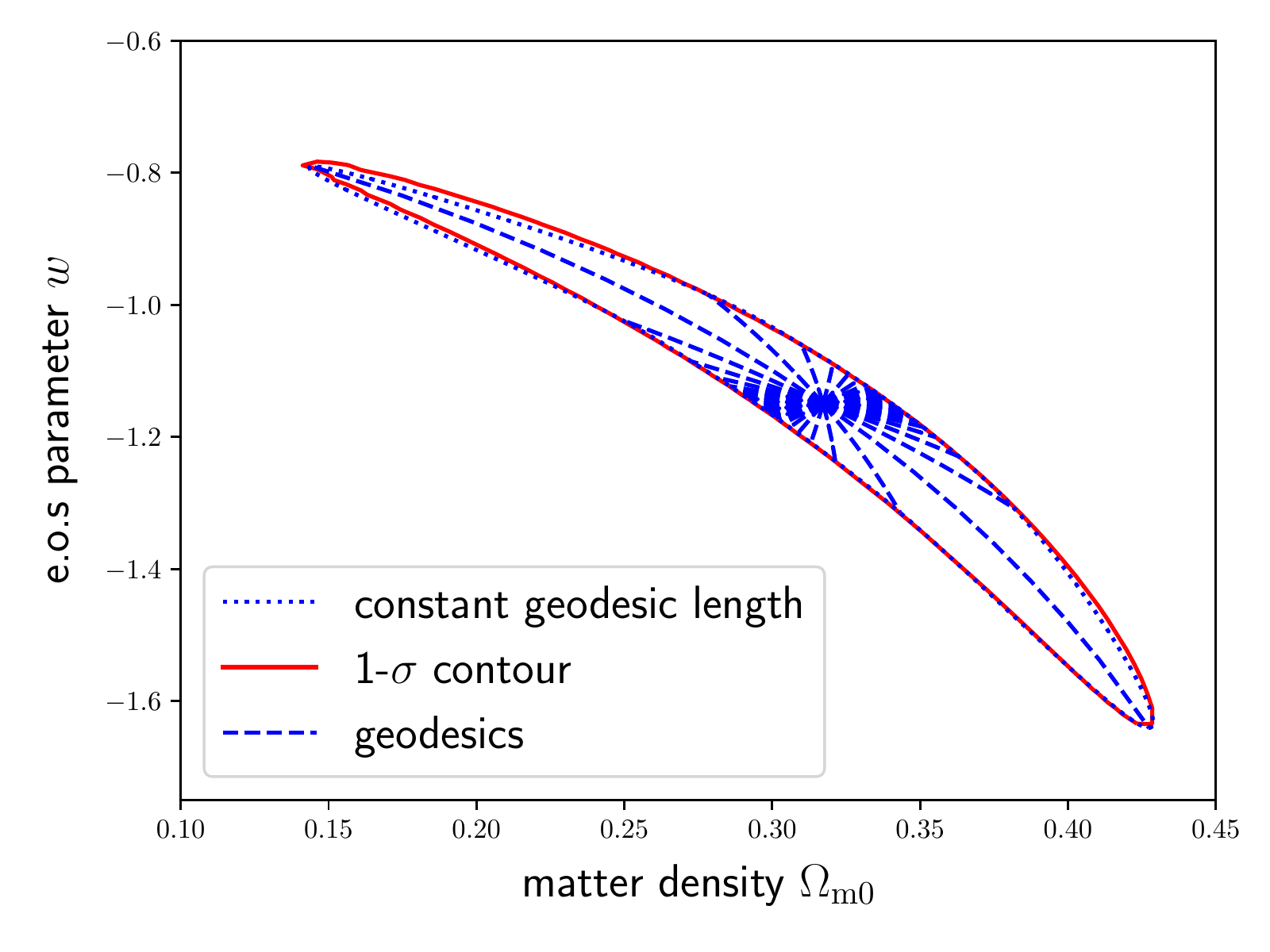}
\caption{\karl{Solutions to the geodesic equation on the statistical manifold of supernovae measurements, where initial velocities are chosen isotropically around the fiducial value in $\Omega_\mathrm{m0}$ and $w$. Dashed blue lines correspond to solutions with different (normalized) initial velocities. The dotted blue line marks the contour of constant geodesic length, $D(P,Q) \approx 2.3$, as motivated from the Gaussian integral in two dimensions. In red we additionally show the 1-$\sigma$ contour from the exact likelihood as in \cref{fig:likelihood}. }}
\label{fig:geodesic}
\end{center}
\end{figure}

\karl{Invariant distances on the manifold are measured as integrals along geodesics, \cref{eq:distance_finite}. These geodesic distances obtain a statistically meaning by looking at a flat manifold, that is a Gaussian. The solution to the geodesic equation are straight lines. If we send out a normalized geodesic into the direction of a single parameter,  one finds the 1-$\sigma$ contour to be defined by the parameter value, were the geodesic distances reaches unity away from the fiducial. This generalizes to more dimensions, resulting into the well known Gaussian integrals. In two dimensions one finds $D(P,Q) \approx 2.3$, which is usually identified as the $\Delta\chi^2$. 
The same probability mass should be enclosed by corresponding contours in the general manifold. It is therefore natural to assume that a geodesic, i.e. a solution to \citep{Nakahara_Geometry_2003}
\begin{equation} 
\ddot{\theta}^a\left(\lambda\right)+\Gamma^{{\mathrm{LC}}\,a}_{bc} \dot{\theta}^b\left(\lambda\right)\dot{\theta}^c\left(\lambda\right)=0,
\end{equation}
sent out in any (normalized) direction from the best fit (or fiducial) value should terminate at the 1-$\sigma$ contour if cut-off at a geodesic length, $D(P,Q) \approx 2.3$.}

\karl{In \cref{fig:geodesic} we show solutions (dashed blue) to the geodesic equation with initial position at the maximum of the exact likelihood. We vary the direction of the initial velocities but always keep them normalized, i.e. $v = (g_{ij}\dot\theta^i\dot\theta^j)|_{\lambda =0} = 1$. Since any metric connection preserves the norm of a vector, $\dot{\boldsymbol{\theta}}$, will stay normalized along the geodesic. Note that any non-normalized vector would just directly turn into a different normalization of $D(P,Q)$. From the solution, we calculate the geodesic length, \cref{eq:distance_finite}, and stop evolving the geodesic equation at $D(P,Q) \approx 2.3$. The resulting contour is shown as a dotted blue line. As a comparison we show the corresponding 1-$\sigma$ contour from the exact likelihood (\cref{fig:likelihood}). Clearly the geodesics trace, as expected, the exact likelihood contour. A few remarks are in order: $(i)$ Although the invariant metric is only determined up to a constant factor, such a redefinition does not change the geodesics, since any metric connection is invariant under any rescaling of the metric. $(ii)$ Geodesics are invariant under affine change of the parametrisation, this changes the geodesic length of the curve. However, a natural parametrisation can always be chosen. Likewise, the tangent vector can always be normalized.} 
\karlo{Lastly, a few comments are in order. Divergences as defined in \cref{eq:divergence_general} are generally not symmetric. They can, however, made symmetric as pointed out before. In general there has been some debate about the validity of information geometry, e.g. \cite{skilling_critique_2014}, since the divergence is not symmetric, while the Riemannian distance clearly is. An alternative approach would be to consider Finsler manifolds instead of a Riemannian manifold. The former is a smooth manifold $M$ equipped with a non-negative function $F$ on its tangent bundle $TM$. Provided $F$ is smooth this object satisfies all the criteria of a divergence.  We plan to investigate this in future work. Furthermore it should be noted that the differences in \cref{fig:geodesic} are mainly due to small numerical uncertainties and should not be interpreted as an effect of the non-symmetry of the divergence or the curvature of the manifold. }

\begin{figure}[h!]
\begin{center}
\includegraphics[keepaspectratio,height=9.7cm, width=9.7cm]{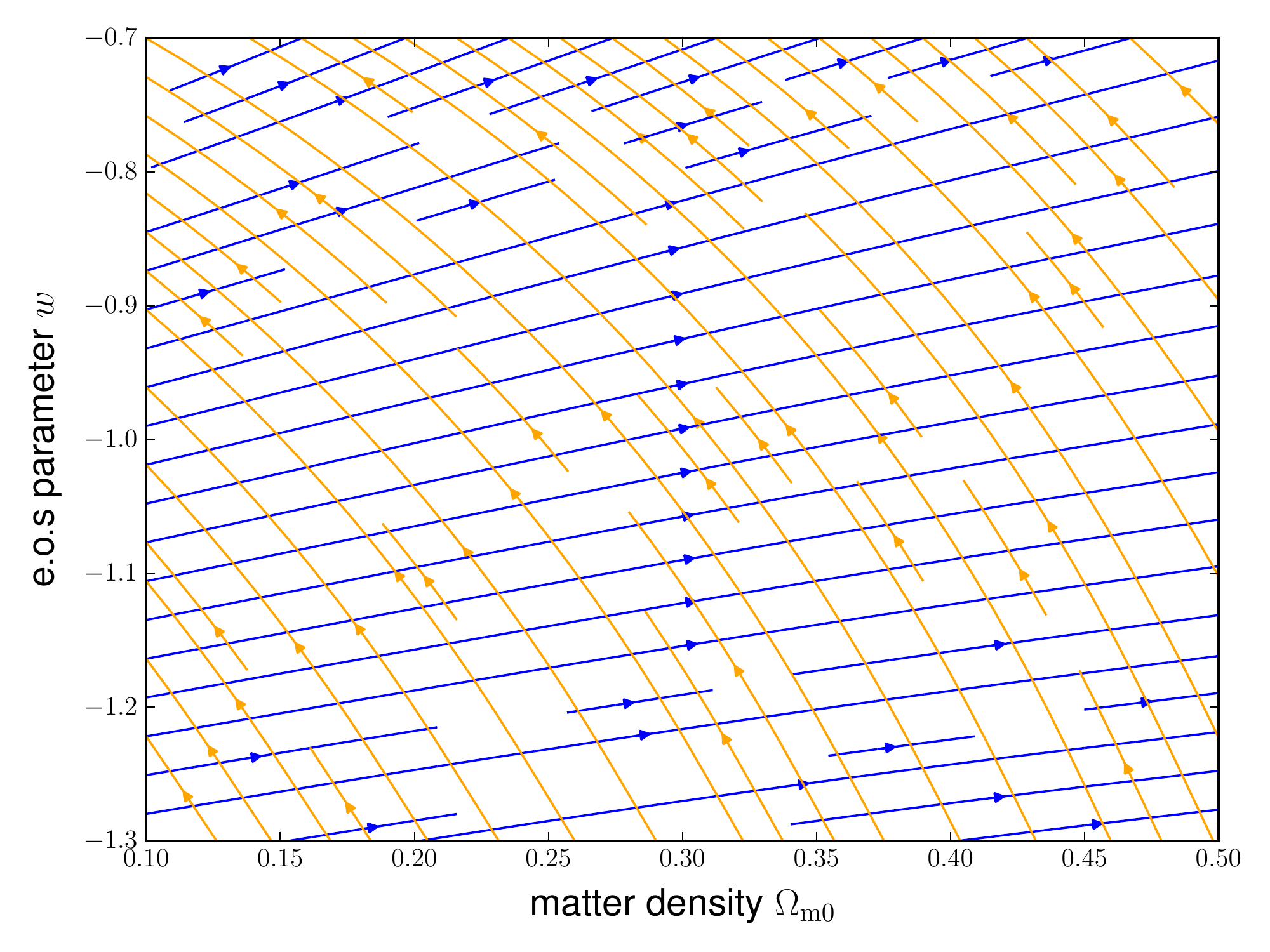}
\caption{\karl{Integral curves of the eigensystem of the Fisher metric of the supernovae likelihood. That is, we show the flow of the vector fields given by the two eigenvectors of the metric $F_{ab}$, \cref{eq:fisher_matrix_sn}. }} 
\label{fig:eigensystems}
\end{center}
\end{figure}

\begin{figure}
\begin{center}
\includegraphics[keepaspectratio,height=9.7cm, width=9.7cm]{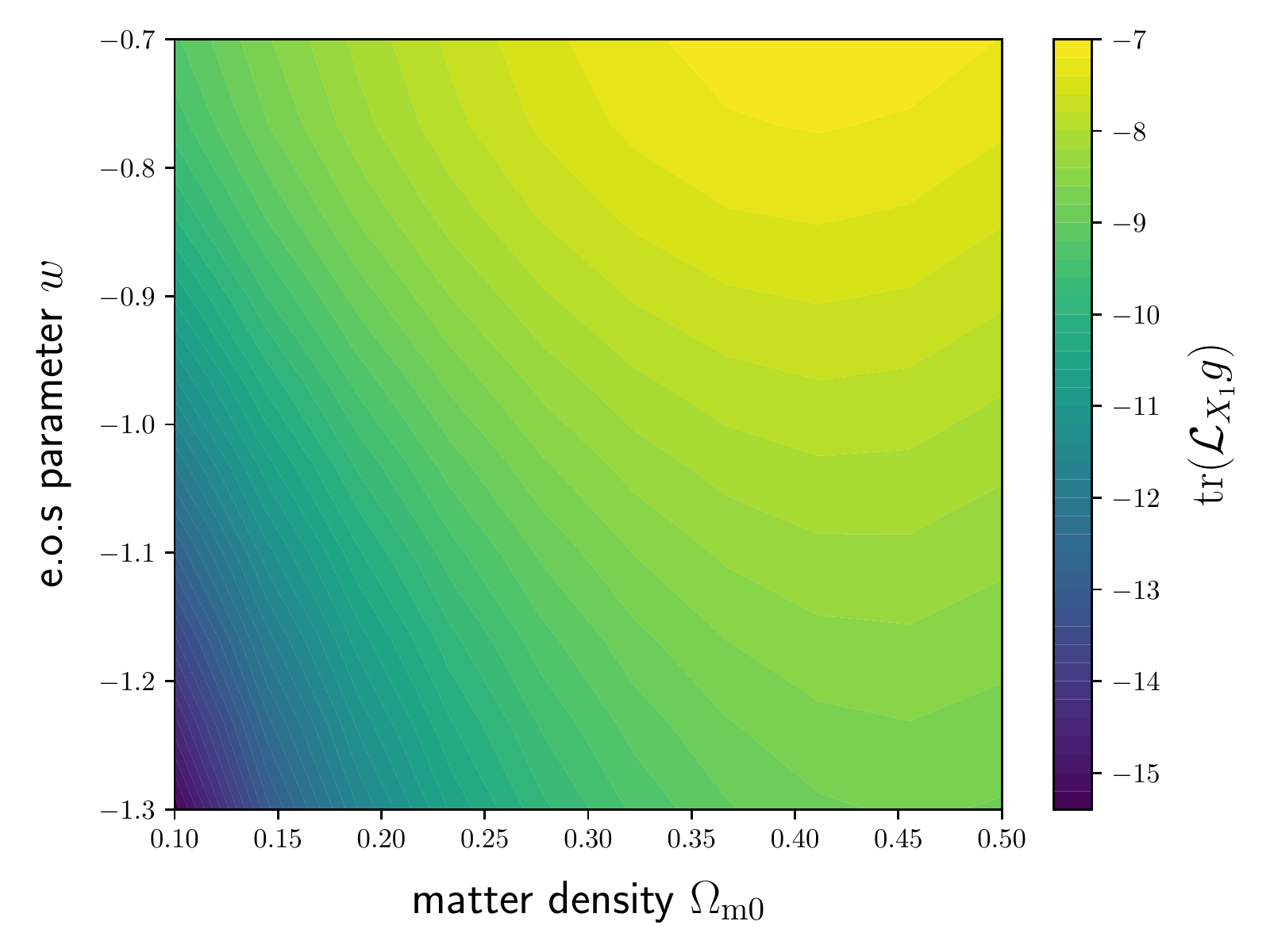}
\includegraphics[keepaspectratio,height=9.7cm, width=9.7cm]{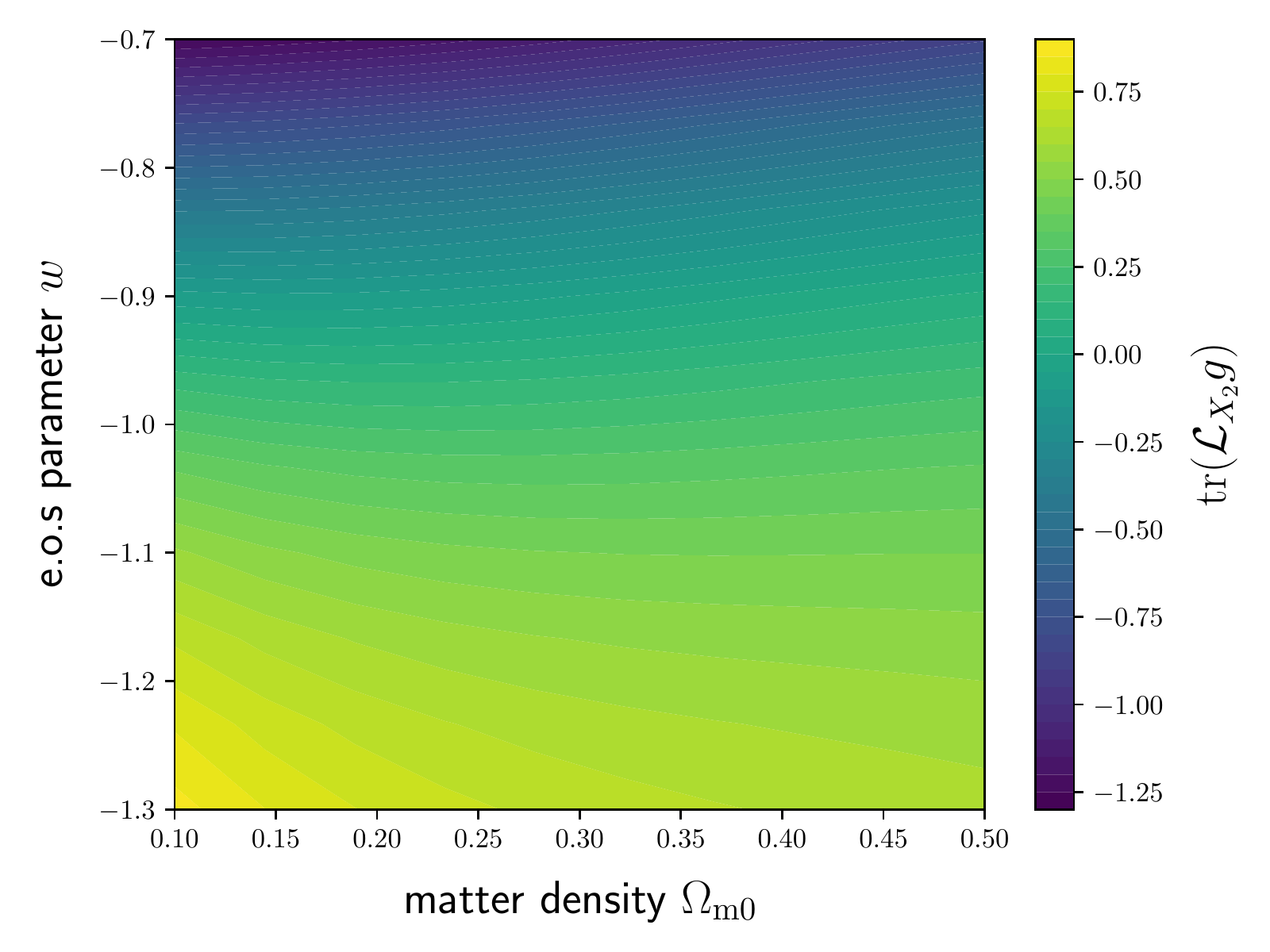}
\caption{\karl{The trace of the Lie derivative, $\mathrm{tr}\left(\mathcal{L}_{\boldsymbol{X}}\boldsymbol{g}\right) = g^{ab}\left(\mathcal{L}_{\boldsymbol{X}}\boldsymbol{g}\right)_{ab}$, along a vector field $\boldsymbol{X}$. The two vector fields under consideration constitute the eigensystem of the Fisher metric. The corresponding flow lines along which the Lie derivative is evaluated is shown in \cref{fig:eigensystems} with blue (top panel) and orange (bottom panel).}}
\label{fig:Lie_derivative}
\end{center}
\end{figure}

\karl{We are now looking for symmetries of the manifold. This is usually done by searching for isometries, which are all those transformations generated by a vector field whose action on the manifold leave the metric invariant. The flow lines of these vector fields are then exactly the isometries. This requires solving the Killing equation, $\mathcal{L}_{\boldsymbol{X}}\boldsymbol{g}=0$ for a vector field $\boldsymbol{X}$, where $\mathcal{L}_{\boldsymbol{X}}\boldsymbol{g}$ is the Lie derivative, with components:
\begin{equation}
\label{eq:lie_derivative_metric}
\left(\mathcal{L}_{\boldsymbol{X}}\boldsymbol{g}\right)_{ab} = 
X_n^c \partial_c g_{ab}+\partial_a X_n^c \,g_{cb}+\partial_b X_n^c \,g_{ac} = 
\partial_a X_b + \partial_b X_a - 2 \Gamma^{\mathrm{LC}\,c}_{ab}X_c.
\end{equation}
Solving the Killing equation for an arbitrary metric is complicated and would require further differentiation to cast it into a finite form. This is beyond the scope of the current paper. Instead we take a different, qualitative approach to illustrate the action of the Lie derivative. 
In \cref{fig:eigensystems} we show the flow of the eigensystem of the metric. The eigensystem itself consist of the normalized vectors $\boldsymbol{X}_1$ and $\boldsymbol{X}_2$ (minor and major axis respectively). We will now use these two vector fields to calculate the Lie-derivative $\mathcal{L}_{\boldsymbol{X}_n}\boldsymbol{g}$ of the metric along them. }

\karl{\Cref{fig:Lie_derivative} shows the trace of the Lie derivative $\mathrm{tr}\left(\mathcal{L}_{\boldsymbol{X}}\boldsymbol{g}\right) = g^{ab}\left(\mathcal{L}_{\boldsymbol{X}}\boldsymbol{g}\right)_{ab}$ as a heat map along the two vector fields (upper and lower panel). One can clearly see that the Lie-derivative with respect to the vector field defined via the minor axis is larger than the one with respect to the major axis. This confirms the observation that the metric changes more strongly along $\boldsymbol{X}_1$. The interpretation of the Lie derivative along these two vector fields is that it quantifies how strongly the geodesic distance changes (equivalently the probability mass between two points on the statistical manifold) if one transforms the manifold along the flow lines of $\boldsymbol{X}_1$. These findings indicate that a Killing vector field of this geometry should be closely aligned with the vector field generated by the major axis of the Fisher matrix. The integral curves of Killing vector fields are symmetry transformations of the metric, i.e. isometries, {meaning that the Lie-derivative of the metric along these vector fields vanishes.} Consequently, they give a direct indications how a non-linear transform should be structured to transform the likelihood into an approximately more Gaussian shape. However, as we have already seen earlier, there is no transformation to make the likelihood completely Gaussian globally.}

% --- section: summary --- %
\section{Summary}\label{sec:summary}
In this paper we have studied a cosmological likelihood and the associated statistical manifold from a differential geometric point of view, using the tools of information geometry, introduced by \citep{amari_information_2016}. After a short description of the methods of information geometry we used them to investigate certain approximations of likelihoods made in cosmology. In particular, we considered weak non-Gaussianities in the Gram-Charlier limit and identified non-Gaussian contributions with the Riemann curvature tensor. Furthermore, we studied the relation between the DALI expansion scheme of likelihoods \citep{sellentin_fast_2015} and showed the connection between its expansion coefficients and geometric objects on the statistical manifold. As an example we studied the likelihood of supernovae in a two-dimensional plane spanned by the two cosmological parameters $\Omega_m$ and $w$. \karl{It should be noted that the methods of information geometry can of course be applied to any non-Gaussian posterior or likelihood, but these non-Gaussian features naturally occur in all branches of cosmology.} Our main findings are the following:

\begin{enumerate}[i)]

\item Local non-Gaussianities in the Gram-Charlier limit can be related to local geometric properties, such as the connection coefficients and the Riemann curvature tensor. 

\item The expansion coefficients of the DALI expansion scheme are directly proportional to higher order cumulants contracted with the Riemannian metric. Furthermore, it can be shown that the flexion, i.e. the first non-trivial expansion coefficient, is related to the cubic tensor. The next order can, however, be shown to vanish in the case of weak non-Gaussianities. 

\item \karl{By applying the tools of information geometry to the supernovae likelihood we could show that it is genuinely non-Gaussian in the $(\Omega_\mathrm{m0},w)$-plane, since the scalar curvature is non-vanishing and a Gaussianisation transform is impossible to construct.}

\item \karl{We studied the relation between geodesic length and the probability mass in a certain region on the statistical manifold. This demonstrated that geodesics with a suitably chosen geodesic length (which can be motivated from the flat manifold) trace out the correct confidence intervals of the exact likelihood. This particular result is not restricted to the case of weak non-Gaussianities but would be applicable to any non-Gaussian unimodal probability distribution.}

\item \karl{As a last, more qualitative application, we further investigated the Lie-derivatives along the degeneracy directions of the Fisher matrices indicating that, if there was an isometry of the Fisher information in this specific example, the respective integral curve should be aligned to the vector field generated by the major axes of the Fisher ellipses. This could lead to an indication of non-linear coordinate transformations to achieve an approximative Gaussianisation, since a global Gaussianisation is not possible.}
\end{enumerate}

For studying the last point further one would have to derive the Killing vector fields numerically through solution of the Killing equation, i.e. find the vector fields for which the Lie-derivative of the metric vanishes identically (if a solution even exists). This could be done by finite differences methods in the parameter domain of interest, choosing appropriate boundary conditions. The partial differential equation can then be expressed in terms of a system of difference equations and solved numerically.

For future studies, further ideas to achieve an approximate Gaussianisation could be to embed the statistical manifold in a higher dimensional Euclidean space \citep{lee_smooth_manifolds}, which however requires the use of hyper-parameters and can become arbitrarily complicated: But examples in statistics exist where extending the parameter space does provide computational advantages. Ultimately, differential geometry ensures obtaining a flat manifold in embedding at the latest when the dimensionality of the embedding is twice as high. Vice versa, one could reduce the statistical likelihood to two dimensions by marginalisation or conditionalisation and then take advantage of the fact that every two-dimensional Riemannian manifold is at least conformally flat \citep{Nakahara_Geometry_2003}. Then, the Fisher metric for a two dimensional statistical manifold could (in principle) be reparametrised to become constant up to a parameter dependent conformal scale factor, and it seems to us that these two avenues are the only ones where a Gaussianisation could be successful, for a genuinely curved statistical manifold. Additionally, we are curious if it is possible to derive that a flat manifold where coordinates can be chosen in a way that the Fisher-information becomes constant, by employing a variational principle: It is a well-known fact that the Shannon-entropy $S = -\int\dd^n\theta\:p(\boldsymbol\theta)\ln p(\boldsymbol\theta)$ is maximised by a Gaussian distribution for a fixed variance, and this result might generalise to implying flatness as a generalisation of the concept of Gaussianity following from variation. Furthermore, a wider class of entropy measures, for instance R{\'e}nyi-entropies $S_\alpha = -\int\dd^n\theta\:p(\boldsymbol\theta)\:p^{\alpha-1}(\boldsymbol\theta)/(\alpha-1)$ for $\alpha\neq 1$, can have interesting geometric implications beyond those of Shannon-entropies $S$.

% --- section: appendix --- %
\appendix

% --- section: multivariate Gram-Charlier series --- %
\section{Remarks on the multivariate Gram-Charlier series}
\label{Multi_Gram}
A way to characterise the properties of a distribution is by its cumulants $\kappa_n$ which are the expansion coefficients of the logarithm of the characteristic function $K\left(t\right) = \ln \tilde{\phi}\left( t\right)= \sum_n \left( \mathrm{i}t \right)^n \kappa_n / n!$ in one dimension \cite{capranico_kalovidaris_schaefer_2013}. The characteristic function $\tilde{\phi}\left(t\right)$ itself is defined as the Fourier transformation of the distribution $p\left(x\right)$. For a multivariate Gaussian $G\left( \boldsymbol{x} \right)$ one can read off the cumulants, which are just the mean $\kappa^{\, \alpha}$ and the covariance $\kappa^{\, \alpha \, \beta}$, from the respective characteristic function $\tilde{\phi}_G\left(\boldsymbol{t}\right)$ as: 
\begin{align}
&G\left( \boldsymbol{x} \right) =  \frac{\sqrt{\det \boldsymbol{C}^{-1}}}{\left(2 \pi \right)^{\,n\,/\,2} } \exp\left( -\frac{\left( x^{\alpha}-\mu^{\alpha} \right)C_{\alpha\beta} \left( x^{\beta}-\mu^{\beta} \right)}{2} \right)
\quad\text{with}\quad
C_{\alpha\beta} \coloneqq \left(C^{\, -1}\right)^{\alpha \, \beta}, 
\
\\
&\tilde{\phi}_G\left(\boldsymbol{t}\right)=\exp\left(\mathrm{i} \, t_{\gamma} \, \mu^{\gamma}  - \frac{ t_{\alpha} C^{\alpha\beta} t_{\beta}}{2} \right)\,\text{with} \,\, \kappa^{\alpha} = \mu^{\alpha} \, \text{as mean and } \kappa^{\alpha\beta} =  C^{\alpha\beta} \, \text{as covariance.}
\end{align}
Here, $\boldsymbol{x}$ and $\boldsymbol{t}$ generalise to a vector in the random variable space and the respective Fourier space. If a distribution has higher order cumulants this is a clear sign of non-Gaussianity. 

For instance one can quantify the asymmetry of a distribution with respect to its peak by the skewness $s \varpropto \kappa_3$, or $\kappa^{\alpha\beta\gamma}$ as multivariate expressions. Furthermore, the kurtosis excess $k \varpropto \kappa_4$, or $\kappa^{\alpha\beta\gamma\delta}$ which characterises the peak morphology is often considered. For $k>0$ the peak appears steeper compared to a Gaussian while for $k<0$ it is flattened. How one can measure these multivariate cumulants is for instance shown in \cite{mardia_measures_1970}. 

The higher order cumulants - beyond the mean and covariance - can now be introduced as small perturbations of a Gaussian characteristic function \cite{capranico_kalovidaris_schaefer_2013}. In the multivariate case this characteristic function with perturbations up to fourth order reads: 
\begin{equation}
\begin{split}
&\tilde{\phi}\left( \boldsymbol{t} \right) = \exp \left[ \frac{\mathrm{i}^3}{3!}t_{\alpha}t_{\beta}t_{\gamma} \, \kappa^{\alpha\beta\gamma} + \frac{\mathrm{i}^4}{4!}\,  t_{\alpha}t_{\beta}t_{\gamma}t_{\delta} \, \kappa^{\alpha\beta \gamma\delta} +\mathcal{O}\left( \boldsymbol{t}^{5} \right)\right] \, \tilde{\phi}_G \left( \boldsymbol{t}  \right),
\end{split}
\end{equation} 
We now perform a Fourier inversion to derive the multivariate Gram-Charlier series. Here the term $ \mathrm{i}^n\, t_{\alpha_1}\ldots t_{\alpha_n} \tilde{f} \left( \boldsymbol{t}  \right)$ is the Fourier transformation of $(-1)^n\, \frac{\partial}{\partial x^{\alpha_1}}\ldots\frac{\partial}{\partial x^{\alpha_n}} \,  f\left( \boldsymbol{x}  \right)$ in complete analogy to the one-dimensional case with $f\left( \boldsymbol{x}  \right)$ being some smooth function. 

The multivariate Gram-Charlier series reads (truncating after fourth order in the cumulants):
\begin{equation}
\begin{split}
p(\boldsymbol{x}) 
&= \exp\left((-1)^3 \frac{\kappa^{\alpha\beta\gamma}}{3!} \frac{\partial^3}{\partial x^{\alpha}\partial  x^{\beta}\partial  x^{\gamma}} +  (-1)^4 \frac{\kappa^{\alpha\beta\gamma\delta}}{4!} \frac{\partial^4}{\partial x^{\alpha}\partial x^{\beta}\partial x^{\gamma}\partial x^{\delta}} \right)
\
\\
& \times \frac{\sqrt{\det \boldsymbol{C}^{-1}}}{\left( 2 \pi \right)^{n/2}}\exp\left(-\frac{\left(x^{\epsilon} - \mu^{\epsilon} \right)C_{\epsilon\zeta}\left(x^{\zeta} - \mu^{\zeta}\right)}{2}\right).
\end{split}
\end{equation}
This expression then simplifies to:
\begin{equation}\label{equ:5.7}
\begin{split}
p(\boldsymbol{x})= \,&\frac{\sqrt{\det\boldsymbol{C}^{-1}}}{\left(2 \pi\right)^{n/2}} 
\exp\left(-\frac{\left(x^{\epsilon} - \mu^{\epsilon} \right)C_{\epsilon\zeta} \,\left(x^{\zeta} - \mu^{\zeta}    \right)}{2}\right)
\left[ 1+ \frac{\kappa^{\lambda\mu\nu}}{3!} \left(W^{-1}\right)_{\lambda\alpha}\left(W^{-1}\right)_{\mu\beta }  \left(W^{-1}\right)_{\nu\gamma  } H_3^{\alpha\beta\gamma} \right.
\
\\
& +\left. \frac{\kappa^{\lambda\mu\nu\xi}}{4!} \left(W^{-1}\right)_{\lambda\alpha} \, \left(W^{-1}\right)_{\mu\beta} \, \left(W^{-1}\right)_{\nu\gamma}\, \left(W^{-1}\right)_{\xi\delta} \,  H_4^{\alpha\beta\gamma\delta}\right].
\end{split}
\end{equation} 
Here, $ H_3^{\alpha\beta\gamma}$ and $H_4^{\alpha\beta\gamma\delta}$ are mulitvariate generalisations of the Hermite polynomials of third and fourth order which are given in \autoref{sec:app_Multivariate_Hermite_Polynomials}.

The covariance matrix can be written as $C^{\alpha\beta} \coloneqq W^{\alpha\gamma} \, W_{\gamma}^{~\beta}$ with $\boldsymbol{W}$ being defined as the matrix root, and respectively $C_{\alpha\beta} \coloneqq \left(C^{\, -1}\right)^{\alpha\beta} = \left(W^{-1}\right)_{\alpha\gamma} \,\left(W^{-1}\right)^{\gamma}_{~\beta}$ for the inverse. The relations (\ref{kumulant_relations}) between the multivariate cumulants and the non-Gaussianities of the DALI-expansion (\ref{eq:dali_expansion}) in \cref{sec:curvature_gaussianization_dali} we can now derive using a similar calculation as for the derivation of the multivariate Gram-Charlier series under the assumption of weak non-Gaussianity. First of all we calculate the characteristic function of the expansion (\ref{eq:dali_expansion}) by employing the same techniques in terms of the Fourier transformation as for the derivation of the multivariate Gram-Charlier series, however now changing from real to Fourier space. Then the result for the characteristic function will also contain an expansion in multivariate Hermite polynomials in the exponent, which have to be written explicitly and compared to the general cumulant expansion of a multivariate characteristic function. Comparison of coefficients finally leads to the relations (\ref{kumulant_relations}).

% --- section: multivariate Hermite polynomials --- %
\section{Multivariate Hermite polynomials}
\label{sec:app_Multivariate_Hermite_Polynomials}
In \autoref{Multi_Gram} a multivariate expression for the Gram-Charlier series is given in equation (\ref{equ:5.7}) which contains multivariate Hermite polynomials. These can be generalised compared to the one-dimensional case as follows:
\begin{equation}\label{equ:B.2}
\begin{split}
H_n^{\, \alpha_1\ldots\alpha_n} = \, & (-1)^n W^{\alpha_1\beta}\ldots W^{\alpha_n\gamma} 
\exp\left(
\frac{\left(x^{\epsilon} - \mu^{\epsilon} \right)\,C_{\epsilon\zeta}\,\left(x^{\zeta} - \mu^{\zeta} \right)}{2}
\right)
\
\\
& \times \frac{\partial}{\partial x^{\beta}}\ldots\frac{\partial}{\partial x^{\gamma} } \, \exp\left(-\frac{\left(x^{\varphi} - \mu^{\varphi} \right)\,C_{\varphi\chi}\,\left(x^{\chi} - \mu^{\chi} \right)}{2}\right).
\end{split}
\end{equation}
Evaluation of relation (\ref{equ:B.2}) up to fourth order will lead to explicit expressions for the multivariate Hermite polynomials:
\begin{equation}
\begin{split}
H_{0} &= 1,
\
\\
H_1^{\,\alpha} &= \left(W^{-1}\right)^{\alpha}_{~\chi} \left(x^{\chi} - \mu^{\chi} \right),
\
\\
H_3^{\, \alpha \, \beta \, \gamma} &= \left(W^{-1}\right)^{\alpha }_{~\lambda} \, \left( x^{\lambda} -\mu^{\lambda} \right)\,\left(W^{-1}\right)^{\beta }_{~\rho} \, \left( x^{\rho} -\mu^{\rho} \right) \, \left(W^{-1}\right)^{\gamma }_{~\tau} \, \left( x^{\tau} -\mu^{\tau} \right) 
\
\\
&\quad - [3]\,  \delta^{\alpha\beta} \left(W^{-1}\right)^{\gamma}_{~\lambda}\left( x^{\lambda} -\mu^{\lambda} \right),
\
\\
H_4^{\alpha\beta\gamma\delta} &= \left(W^{-1}\right)^{\alpha }_{~\lambda}  \left( x^{\lambda} -\mu^{\lambda} \right)\left(W^{-1}\right)^{\beta }_{~\rho}  \left( x^{\rho} -\mu^{\rho} \right)  \left(W^{-1}\right)^{\gamma }_{~\tau}  \left( x^{\tau} -\mu^{\tau} \right)
\left(W^{-1}\right)^{\delta }_{~\eta}  \left( x^{\eta} -\mu^{\eta} \right)
\
\\
&\quad -[6]\, \delta^{\alpha\beta} \left(W^{-1}\right)^{\gamma}_{~\lambda} \left(W^{-1}\right)^{\delta}_{~\eta}\left( x^{\lambda} -\mu^{\lambda} \right)\left( x^{\eta} -\mu^{\eta} \right
)+[3]\, \delta^{\alpha\beta} \, \delta^{\gamma\delta}.
\end{split}
\end{equation}
Here the short-hand notation $\left[n\right]$ means that $n$ terms with permutation in indices exist, while $\delta^{\alpha\beta}$ denotes the Kronecker-Symbol.
%\label{sec:app_differential_geometry}

% --- bibliography --- %
\bibliographystyle{JHEP}
\bibliography{references,My_Library}

% --- section: acknowledgements --- %
\acknowledgments
RR acknowledges funding through the HEiKA-initative and support by the Israel Science Foundation (grant no. 1395/16 and grant no. 255/18). EG thanks the Studienstiftung des deutschen Volkes and CERN's Wolfgang-Gentner Programme for financial support. The authors thank an anonymous referee for improving the manuscript substantially. We also thank Marie Teich, Rafael Arutjunjan and Nils Fischer for insightful discussion, questions and remarks.
\end{document}